\RequirePackage{fix-cm}
\documentclass[smallextended]{svjour3}       
\smartqed  
\usepackage{graphicx}
\usepackage{amsmath}
\usepackage{amssymb}
\usepackage{xcolor}
\begin{document}

\title{Local two- and three-nucleon chiral interactions
}

\author{Maria Piarulli         \and
        Rocco Schiavilla 
}


\institute{Maria Piarulli \at
              Washington University in St. Louis \\
              \email{m.piarulli@physics.wustl.edu}           
           \and
           Rocco Schiavilla \at
              Old Dominion University and Jefferson Lab\\
               \email{schiavil@jlab.org}  
}

\date{Received: date / Accepted: date}

\maketitle

\begin{abstract}
Understanding the structure and reactions of nuclei from first principles has been a long-standing goal of nuclear
physics. In this respect, few- and many-body systems provide a unique laboratory for studying nuclear interactions.
In the past couple of decades, the modeling of nuclear interactions has progressed significantly owing, in particular,
to the development of chiral effective field theory ($\chi$EFT), a low-energy effective representation of quantum
chromodynamics (QCD).  Within $\chi$EFT, many studies have dealt with the construction of
both two- and three-nucleon interactions.  The aim of the present article is to
provide a concise account of chiral interaction models that are local in configuration space, and
to report on a selection of recent results for nuclear systems obtained with these interactions.
\end{abstract}

\section{Introduction}
The modeling of nuclei as systems of nucleons (protons and
neutrons) interacting with each other via effective forces and with
external electroweak probes via effective currents has a long
and venerable history.  We refer to it as the basic model of nuclear
physics.  When combined with accurate methods to solve the
many-body Schr\"{o}dinger equation, the basic model presents
us with the opportunity and challenge to understand and explain
nuclear structure and reactions in terms of the underlying dynamics
of interacting nucleons.  A calculation carried out in such a framework
is commonly referred to as an {\it ab-initio} one.  Examples of {\it ab initio}
calculations are
those based on the no-core shell model (NCSM)~\cite{Barrett:2013nh,Jurgenson:2013yya},
the coupled cluster (CC)~\cite{Hagen:2013yba,Hagen:2013nca} or
hyperspherical harmonics (HH)~\cite{Kievsky:2008es} expansions,
similarity renormalization group (SRG) approaches~\cite{Bogner:2009bt,Hergert:2012nb},
self-consistent Green's function techniques~\cite{Dickhoff:2004xx,Soma:2012zd},
quantum Monte Carlo (QMC) methods~\cite{Carlson:2014vla,Gandolfi:2020pbj},
and nuclear lattice effective field theory (NLEFT)~\cite{Lahde:2013uqa}. 
While significant progress has been made in recent years, enabled by advances in the input nuclear interactions and currents based on chiral effective field theory ($\chi$EFT), improved and novel many-body frameworks, and increasingly powerful computer facilities, these {\it ab initio}
calculations remain challenging and their domain of applicability is, at present,
limited to provide a quantitative description of light and medium-mass nuclei~\cite{Barrett:2013nh,Hagen:2013nca,Hergert:2012nb,Soma:2012zd,Dickhoff:2004xx,Carlson:2014vla,Gandolfi:2020pbj,Hagen:2015yea} and their reactions~\cite{Lovato:2014eva,Lovato:2019fiw,Hupin:2014iqa,Elhatisari:2015iga}. The main challenge is to describe diverse physical phenomena within a single coherent picture. The reasons are twofold.
First, at the moment, there exist no interactions and electroweak currents which are able to correctly predict, simultaneously, different nuclear few- and many-body observables over a wide range of mass number, including infinite matter, within quantified theoretical uncertainties. This can be probably traced back to fundamental questions regarding
inconsistencies in the derivation and implementation of nuclear interactions and current operators, and the complexity of the optimization procedures needed for estimating the parameters entering the nuclear models. 
Second, the difficulty in the solution of the nuclear many-body problem is exacerbated by limitations inherent
to the different many-body frameworks utilized for atomic nuclei and nuclear matter. These drawbacks include the scaling behavior as a function of mass number, the convergence of observables as a function of basis states, the validity of many-body truncations, and constraints regarding which nuclear interactions can be used. A special but related challenge is the development, within the basic model,
of approaches accounting for the coupling to the continuum---these are mandatory to describe, for instance, weakly bound nuclear systems~\cite{Fossez:2017wpa,Fossez:2018gae}.\\
\indent Of course, nucleons are composite particles, and it could be argued that an
understanding of nuclei that is truly fundamental can only be realized on
the basis of approaches explicitly (as opposed to effectively) accounting for the
dynamics of quarks and gluons, the degrees of freedom of Quantum
Chromodynamics (QCD).  Such approaches, which are computationally
very demanding, attempt to solve the nuclear many-body problem on a
discretized (Euclidean) space-time lattice.  Albeit there have been many
advances~\cite{Inoue:2013nfe,Beane:2014ora,Orginos:2015aya,Savage:2016kon},
lattice QCD calculations are still limited to small nucleon numbers and/or large pion masses,
and hence, at the present time, can only be used to address a limited set of
representative key-issues.  As a consequence, most theoretical studies of
nuclear systems must turn to the basic model to address the
full complexity of the nuclear many-body problem.
\section{Nuclear interactions}
The basic model assumes that a Hamiltonian consisting of non-relativistic kinetic energy, and
two-nucleon ($2N$) and three-nucleon ($3N$) interaction, provides a good approximation to
the energy of interacting nucleons. 

Two-nucleon interactions are characterized by a long-range component, due to one-pion
exchange (OPE)~\cite{Yukawa:1935xg}, for inter-nucleon separation $r \gtrsim 2$ fm, and
intermediate- and short-range components for $1 \,\,{\rm fm}  \lesssim r  \lesssim 2$ fm and
$r \lesssim 1$ fm, respectively.  Up until the mid-1990's, these
interactions~\cite{Stoks:1994wp,Wiringa:1994wb,Machleidt:2000ge} were based essentially
on meson-exchange phenomenology, with parameters characterizing the short- and
intermediate-range components that were constrained by fits to the $2N$ elastic scattering
data up to lab energies of 350 MeV (that is, slightly above the threshold for pion production).
The $\chi^2$/datum achieved in these fits was close to 1 relative to the database available
at the time~\cite{Stoks:1993tb}.  Two well-known, and still widely used, examples in this class
of {\sl phenomenological} $2N$ interactions are the Argonne $v_{18}$ (AV18)~\cite{Wiringa:1994wb}
and CD-Bonn~\cite{Machleidt:2000ge}. 

Already in the early 1980's, accurate Faddeev calculations had shown that $2N$ interactions (those
available at the time) did not provide enough binding for the three-body nuclei, $^3$H and
$^3$He~\cite{Friar:1984ic}.  In the late 1990's and early 2000's this conclusion was shown
to hold also for the energy spectra (ground and low-lying excited states) of light p-shell nuclei 
in calculations based on the phenomenological interactions mentioned earlier, and using quantum Monte
Carlo (QMC)~\cite{Pudliner:1997ck} and no-core shell-model (NCSM)~\cite{Navratil:2000gs} methods.
This led to the realization that the basic model without the inclusion of (at least) $3N$ interaction is definitely
incomplete.

Because of the composite nature of the nucleon and, in particular, the
dominant role of the $\Delta$-resonance in pion-nucleon scattering,
multi-nucleon interactions arise quite naturally in the meson-exchange
phenomenology.  In particular, the Illinois $3N$ interactions~\cite{Pieper:2001ap}
consist of a dominant two-pion exchange (TPE) component with a single
intermediate $\Delta$---the Fujita-Miyazawa interaction~\cite{Fujita:1957zz}---and
smaller multi-pion exchange components resulting from the excitation of multiple
intermediate $\Delta$'s. The most recent version, Illinois-7 (IL7)~\cite{Pieper:2008rui},
also contains phenomenological isospin-dependent central terms. The few (4) parameters
characterizing the IL7 model have been determined by fitting the low-lying spectra of nuclei
in the mass range $A\,$=$\,$3--10. The resulting AV18+IL7 Hamiltonian, generally utilized
with QMC methods, then leads to predictions of about 100 ground- and excited-state energies
up to $A\,$=$\,$12, including the $^{12}$C ground- and Hoyle-state energies, in good agreement
with the corresponding empirical values~\cite{Carlson:2014vla}.  However, when used to compute
the equation of state of neutron star matter, the AV18+IL7 Hamiltonian does not provide sufficient
repulsion to ensure the stability of the observed stars against gravitational collapse~\cite{Maris:2013rgq}.
Thus, it would appear that, in the context of phenomenological nuclear interactions, we do not have
models that can predict simultaneously the properties of light p-shell nuclei
and dense nuclear and neutron matter.  It is also important to emphasize
that these interactions are affected by several additional limitations, most notably
the missing link with the (approximate) chiral symmetry exhibited by QCD,
and the absence of rigorous schemes to consistently derive nuclear electroweak currents.

The advent of chiral effective field theory ($\chi$EFT)~\cite{Weinberg:1990rz,Weinberg:1991um,Weinberg:1992yk}
in the early 1990's has spurred a new phase in the evolution of the basic model, and has renewed interest
in its further development. $\chi$EFT is a low-energy effective theory of QCD based on pions and
nucleons (and, in some instances, $\Delta$'s) as effective degrees of freedom.  For momenta $p\sim m_{\pi}$,
such a framework is expected to be accurate, since shorter-range structures, e.g., the quark substructure, or
heavier meson exchanges, e.g., $\rho$-meson exchanges, are not resolved, and can be absorbed in
short-range contact interactions between nucleons.  This {\it separation of scales} between typical
momenta $p\sim m_{\pi}$ and much harder momenta of the order of the $\rho$-meson or nucleon mass
can be used to systematically derive a general scheme, which accommodates all possible
interactions among the relevant degrees of freedom (pions, nucleons, and, in some formulations, $\Delta$'s), 
and which is consistent with the symmetries of QCD. 

The starting point in $\chi$EFT is the most general Lagrangian in terms of the
chosen degrees of freedom, which contains all interaction mechanisms allowed
by the symmetries of QCD.  This Lagrangian contains an infinite number of terms
and needs to be truncated using a given power-counting scheme.  Most chiral
interactions used in nuclear structure calculations use Weinberg's power counting,
which itself is based on naive dimensional analysis of interaction contributions. 
Within Weinberg's power counting, the interactions are expanded in powers of the
typical momentum $p$ over the breakdown scale $\Lambda_b$, that is, the
expansion parameter is $Q=p/\Lambda_b$, where the breakdown scale denotes
momenta at which the short distance structure becomes important and cannot be
neglected and absorbed into contact interactions any longer
(see Refs.~\cite{Epelbaum:2008ga,Machleidt:2011zz,Machleidt:2016rvv,Machleidt:2017vls,Hammer:2019poc}
for recent review articles).  It is worthwhile mentioning that alternative power-counting
schemes have been also suggested, see Refs.~\cite{Kaplan:1998tg,Kaplan:1998we,Nogga:2005hy,PavonValderrama:2005wv,Long:2011xw,vanKolck:1994yi}.

This expansion defines an order by order scheme, defined by the power $\nu$ of the
expansion parameter $Q$ in each interaction contribution: leading order (LO) for $\nu=0$,
next-to-leading order (NLO) for $\nu=2$, next-to-next-to-leading order (N$^2$LO) for $\nu=3$
and so on. Similarly as for nuclear interactions, such a scheme can also be developed for
electroweak currents~\cite{Rho:1991}.  Therefore, $\chi$EFT provides a rigorous scheme to
systematically construct nuclear many-body forces and consistent electroweak currents,
and tools to estimate their
uncertainties~\cite{Furnstahl:2014xsa,Epelbaum:2014efa,Furnstahl:2015rha,Wesolowski:2015fqa,Melendez:2017phj,Wesolowski:2018lzj}. 

Nuclear interactions in $\chi$EFT are separated into pion-exchange terms, associated with
the long- and intermediate-range components, and contact terms that encode short-range
physics.  The strength of these contact terms is specified by unknown low-energy constants
(LECs), which are constrained by fitting experimental data.  Nuclear interactions (and
electroweak currents) suffer from ultraviolet (UV) divergencies, which need to be removed
by a proper regularization and renormalization procedure.  As a matter of fact, there are two
sources of UV divergencies that require regularization: one from loop corrections and
the other when solving the Schr\"{o}dinger equation (or when calculating
matrix elements of nuclear currents).  Loop divergences can be treated via dimensional
regularization (DR) or spectral-function regularization (SFR), where the latter is implemented
by including a finite cutoff in the spectral functions. If this cutoff is taken to be infinity, then
SFR coincides with DR.  To remove divergencies occurring in the solution
of the Schr\"{o}dinger equation, nuclear interactions are multiplied by regulator
functions that remove momenta larger than a preset cutoff scale.  The regularization
of interactions (and currents) is followed by a renormalization procedure, that is,
dependencies on the regularization scheme and cutoff are reabsorbed, order by
order, by the LECs characterizing these interactions (and currents).

Nucleon-nucleon scattering has been extensively studied in $\chi$EFT in the past
two decades following the pioneering work by Weinberg~\cite{Weinberg:1990rz,Weinberg:1991um,Weinberg:1992yk}
and Ordonez {\it et al.}~\cite{Ordonez:1995rz}.   In particular, $2N$ interactions at N3LO
in the chiral expansion are available since the early 2000's~\cite{Entem:2003ft,Epelbaum:2004fk}
and have served as a basis for numerous {\it ab initio} calculations of nuclear structure and reactions.
More recently, models up to the fifth order in the chiral expansion, i.e., N4LO,
have been developed~\cite{Entem:2015xwa,Epelbaum:2014sza,Reinert:2017usi,Entem:2017gor}, which
lead to accurate descriptions of $2N$ databases up to laboratory energies of
300 MeV with $\chi^2$ per datum close to 1. 
These databases have been provided by the Nijmegen group~\cite{Stoks:1993tb,Stoks:1994wp}, the VPI/GWU group~\cite{Arndt:2007qn}, and more recently the Granada group~\cite{Perez:2013jpa,Perez:2013oba,Perez:2014yla}. 
In the standard optimization procedure, the $2N$ interactions are first constrained
through fits to neutron-proton ($np$) and proton-proton ($pp$) phase shifts,
and then refined by minimizing the total $\chi^2$ obtained from a direct comparison
with the $2N$ scattering data.  However, new optimization schemes are being
explored in Refs.~\cite{Carlsson:2015vda,Ekstrom:2015rta}.
For instance, the optimization strategy of Ref.~\cite{Ekstrom:2015rta} is based
on a simultaneous fit of low-energy $2N$ data,
the deuteron binding energy, and the binding energies and charge radii of hydrogen,
helium, carbon, and oxygen isotopes using consistent $2N$ {\it and} $3N$ interactions at N2LO. 


Three-nucleon interactions and their impact on nuclear structure and reactions have
become a nuclear-physics topic of intense current interest, see Refs.~\cite{KalantarNayestanaki:2011wz,Hammer:2012id,Hebeler:2020ocj}
for review articles.  Three-nucleon interactions have been derived up to N4LO in
$\chi$EFT~\cite{Bernard:2007sp,Bernard:2011zr,Krebs:2012yv,Krebs:2013kha,Girlanda:2011fh}. 
However, few- and many-nucleon calculations are, with very few exceptions, still limited to chiral $3N$
interactions at N2LO. At this order, they are characterized by two unknown LECs, one 
in a OPE-contact term and the other in a purely contact $3N$ term; these LECs are commonly denoted
as $c_D$ and $c_E$, respectively.  They have been constrained either by fitting exclusively strong-interaction observables~\cite{Tews:2015ufa,Lynn:2015jua,Lynn:2017fxg,Piarulli:2017dwd} or by relying on
a combination of strong- and weak-interaction observables~\cite{Gazit:2008ma,Marcucci:2011jm,Baroni:2018fdn,Schiavilla:2017,Wesolowski:2021cni}.
This last approach is made possible by the relation between $c_D$ and the LEC entering the $2N$ contact axial 
current~\cite{Gardestig:2006hj,Gazit:2008ma,Marcucci:2011jm}.  This relation
emerges naturally in $\chi$EFT, and allows one to use nuclear properties governed by either strong
or weak interactions to constrain simultaneously the $3N$ interaction and $2N$ axial current.

Since $\chi$EFT is a low-momentum expansion, many of the chiral interactions
available in the literature are naturally formulated in momentum space and have the feature of being strongly
non-local in coordinate space.  This makes them ill-suited for certain numerical algorithms, for example,
Quantum Monte Carlo (QMC) methods. This strong non-locality comes about on account of
two factors: (i) the specific choice made for the cutoff function needed to remove large momenta, and
(ii) contact terms involving high-order derivatives of the nucleon field.

\section{Local Chiral interactions}
In recent years, local chiral interactions suitable for QMC calculations have been developed by two
different groups using 
$\Delta$-less~\cite{Gezerlis:2013ipa,Gezerlis:2014zia,Lynn:2015jua,Lynn:2017fxg,Lonardoni:2018nob,Lynn:2019rdt}
and $\Delta$-full~\cite{Piarulli:2016vel,Piarulli:2014bda,Baroni:2018fdn,Piarulli:2017dwd,Piarulli:2020mop}
$\chi$EFT formulations.  At LO, both $\Delta$-less and $\Delta$-full interactions have the same
operator structure.  At this order, only the leading contact terms (involving no derivatives
of the nucleon field) and one-pion exchange (OPE) term contribute (the latter is often taken
to include also the charge-independence breaking induced by the difference between the
neutral and charged pion masses). 

At higher orders, additional momentum-dependent contact as well as two-pion
exchange (TPE) terms appear.  The TPE coordinate-space expressions at
NLO and N2LO for both the $\Delta$-less and $\Delta$-full approaches are given in Refs.~\cite{Epelbaum:2003gr,Gezerlis:2013ipa,Gezerlis:2014zia} and Ref.~\cite{Piarulli:2014bda},
respectively.  For the NLO contact interactions, the most general form consists of 14 terms~\cite{Machleidt:2011zz}. However, only 7 out of these 14 terms are linearly
independent; they turn out to be fully local.  Moreover, at this order, a leading contact charge-dependent
(CD) term is also accounted for, needed to reproduce the $pp$ and $nn$ singlet scattering
length.  

At the next order, N3LO, contact interactions cannot be written down in a purely local fashion,
since Fierz identities prove ineffective in removing all non-localities. A possible way forward
is the definition of {\it minimally non-local} N3LO interactions, which have been constructed in the
$\Delta$-full approach as reported in Ref.~\cite{Piarulli:2014bda}. The local versions of these
$\Delta$-full {\it minimally non-local} $2N$ interactions have been defined by dropping terms
proportional to ${\bf p}^2$ that remain after Fierz rearrangement~\cite{Piarulli:2016vel} (here,
${\bf p}$ is the relative momentum operator). The inclusion of these terms was shown to yield
no significant improvement in the fit to the $2N$ database~\cite{Piarulli:2016vel}. As a matter of fact,
three combinations of such terms vanish off the energy shell~\cite{Reinert:2017usi} and
their effect can be absorbed into a redefinition of the $3N$ interaction~\cite{Girlanda:2020pqn}.
In these models, four charge-dependent (CD) operators at N3LO are also retained~\cite{Piarulli:2016vel}.

In order to use these interaction models in many-body calculations, it is necessary to specify a
regularization scheme. For the $\Delta$-less interactions, the following long- and short-range
regulators are used~\cite{Gezerlis:2013ipa,Gezerlis:2014zia},
\begin{align}
    f_{\rm{long}}(r)&=\left[ 1- e^{ -(r/R_0)^{n_1} } \right]^{n_2} \,,\quad 
    f_{\rm{short}}(r)=\frac{n}{4 \pi\,R_0^3\,\Gamma\left(3/n\right)}\,e^{-\left(r/R_0 \right)^{n}} \,, \label{eq:fshort}
\end{align} 
with $n_1=4$, $n_2=1$, and $n=4$.

The long-range regulator multiplies each radial function in the OPE and TPE contributions,
while the short-range regulator replaces all $\delta$-functions in the contact terms. The
regulator functions depend on the cutoff scale $R_0$ that is taken in the range of $R_0$ = (1.0--1.2) fm.
There are 11 LECs associated with contact terms in the $\Delta$-less (NLO) models.  They are fixed by performing
$\chi^2$ fits to $2N$ phase shifts from the Nijmegen partial-wave analysis (PWA) up to 150 MeV laboratory energy~\cite{Gezerlis:2013ipa,Gezerlis:2014zia}.

In the $\Delta$-full interactions, the long- and short-range regulators are, instead, given by the following functions 
\begin{align}
    f_{\rm long}^{\Delta}(r)=1-\frac{1}{(r/R_{\rm L})^6 \,  e^{(r-R_{\rm L})/a_{\rm L}} +1},\quad 
    f_{\rm short}^{\Delta}(r)=\frac{1}{\pi^{3/2}R_{\rm S}^3} e^{-(r/R_{\rm S})^2} \, ,
\end{align} 
where three values for the radius $R_{\rm L}$ are considered: $R_{\rm L}=(0.8,1.0,1.2)$ fm with the diffuseness $a_{\rm L}$ fixed at $a_{\rm L}=R_{\rm L}/2$ in each case. In combination with $R_{\rm L}$, the $R_{\rm S}$ values considered
are $(0.6,0.7,0.8)$ fm, corresponding to typical momentum-space cutoffs $\Lambda_{\rm S}=2/R_{\rm S}$ ranging from about 660 MeV down to 500 MeV.  The interactions with cutoffs $(R_{\rm L},R_{\rm S})$ equal to
$(1.2,0.8)$ fm, $(1.0,0.7)$ fm, and  $(0.8,0.6)$ fm are denoted, respectively, as model a, b, and c.
There are 26 LECs that enter these (N$^2$LO) interactions. The optimization procedure to fix these 26 LECs
utilizes $pp$ and $np$ scattering data (including normalizations), as assembled in the Granada database~\cite{Perez:2013jpa}, the $2N$ scattering lengths, and the deuteron binding energy. 
For each of the three different sets of cutoff radii $(R_{\rm S},R_{\rm L})$, two classes of local
interactions have been developed, which only differ in the range of laboratory energy over which
the fits were carried out, either 0--125 MeV in class I or 0--200 MeV in class II. 
The $\chi^2$/datum achieved by the fits in class I (II) was $\lesssim 1.1(\lesssim1.4)$
for a total of about 2700 (3700) data points.  In the literature, these $2N$ interactions are
generically referred to as the Norfolk interactions (NV2s).  Those in class I are designated
as NV2-Ia, NV2-Ib, and NV2-Ic, and those in class II as NV2-IIa, NV2-IIb, and NV2-IIc. 

Both the $\Delta$-less and $\Delta$-full formulations account for $3N$ interactions. 
In the $\Delta$-less version, the leading $3N$ contributions appear at N2LO in the
power counting.  They consist of (i) a long-range TPE term ($V_C$), 
depending on the subleading pion-nucleon LECs $c_1$, $c_3$, and $c_4$, that already appear
in the $2N$ sector; (ii) a OPE-contact term ($V_D$) dependent on the LEC $c_D$, and (iii) a purely
 contact $3N$ term ($V_E$) dependent on the LEC $c_E$.
 The LECs $c_D$ and $c_E$ are adjusted so as to fit properties of $A \geq 3$ systems.
 In the $\Delta$-less approach, these observables have been chosen
 to be the $^4$He binding energy and $n$-$\alpha$ scattering $P$
 wave phase shifts. In Fig.~1 of Ref.~\cite{Lynn:2015jua}, the parameter curves for the $3N$ LECs
 corresponding to different $3N$ cutoffs $R_{\text{3N}}$, chosen similarly to $R_0$, are shown.

In the $\Delta$-full formulation, the $3N$ interaction consists of the three N2LO terms
above ($V_{C}$, $V_{D}$ and $V_{E}$) plus a NLO TPE term involving the excitation
of a $\Delta$ in the intermediate state, the well-known
Fujita-Miyazawa interaction~\cite{Fujita:1957zz} ($V_{\Delta}$).
In the $\Delta$-less approach, it is expected to be subsumed in
$V_C$. In the $\Delta$-full chiral EFT, two different sets for the values of $c_D$ and $c_E$
were obtained, leading to two different parametrizations of the $3N$
interaction~\cite{Piarulli:2017dwd,Baroni:2018fdn}. In the first, these LECs were determined
by simultaneously reproducing the experimental trinucleon ground-state energies and the
neutron-deuteron ($nd$) doublet scattering length, as shown in Ref.~\cite{Piarulli:2017dwd}.
In the second set, these $c_D$ and $c_E$ were constrained by fitting, in addition to the
trinucleon energies, the empirical value of the Gamow-Teller matrix element in tritium $\beta$ decay~\cite{Baroni:2018fdn}. Because of the much reduced correlation between binding energies
and the GT matrix element, the second procedure leads to a more robust determination of
$c_D$ and $c_E$ than attained in the first one. Note that these observables have been
calculated with hyperspherical-harmonics (HH) expansion methods~\cite{Kievsky:2008es} as described in Refs.~\cite{Piarulli:2017dwd,Baroni:2018fdn}.

\section{Applications}
In this section, we briefly discuss some illustrative applications of local chiral interactions to the few- and many-body systems.
\begin{figure}[t]
\centering
\includegraphics[width=4.5in]{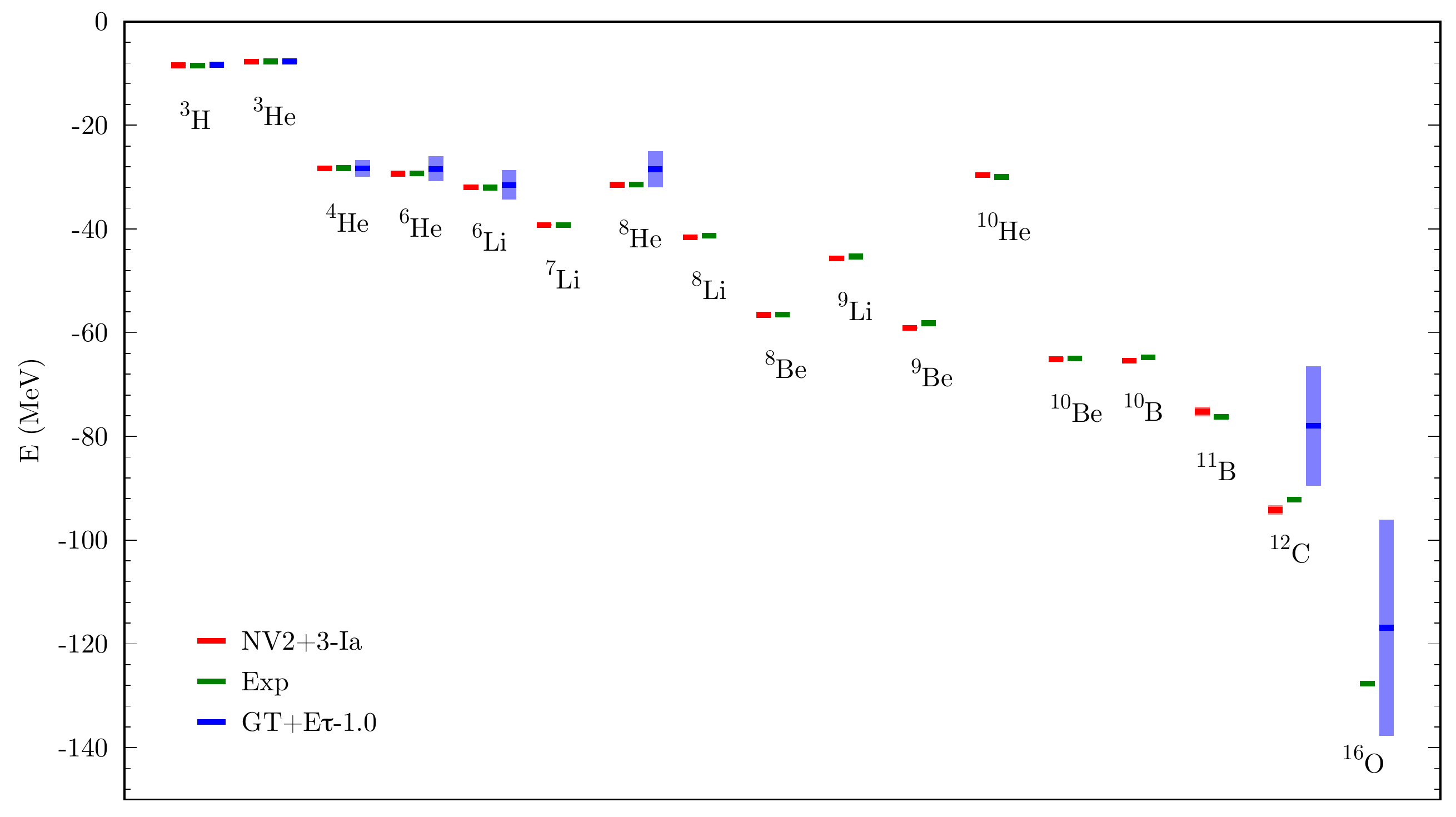}
\caption{\label{fig:NucleiResults}
From Ref.~\cite{Gandolfi:2020pbj}. Ground-state energies in $A\le16$ nuclei. For each nucleus, experimental results~\cite{Wang:2017} are shown in green at the center. GFMC (AFDMC) results for the NV2+3-Ia~\cite{Piarulli:2017dwd} (GT+E$\tau$-1.0~\cite{Lonardoni:2018nob}) interactions are shown in red (blue) to the left (right) of the experimental values. For the NV2+3-Ia (GT+E$\tau$-1.0) interactions, the colored bands include statistical (statistical plus systematic) uncertainties.}
\end{figure} 

Figure~\ref{fig:NucleiResults} shows the binding energies of nuclei up to $^{16}$O as calculated with the Green's function Monte Carlo (GFMC) method for one of the $\Delta$-full models (NV2+3-Ia)~\cite{Piarulli:2017dwd}, and with the Auxiliary Diffusion Monte Carlo (AFDMC) method for one of the $\Delta$-less models (GT+E$\tau$-1.0)~\cite{Lonardoni:2017hgs,Lonardoni:2018nob}. The calculated energies are compared to the experimental values. GFMC results only carry Monte Carlo statistical uncertainties, while for AFDMC results, theoretical uncertainties coming from the truncation of the chiral expansion are also included. These uncertainties are estimated accordingly to the prescription of Epelbaum \textit{et al.}~\cite{Epelbaum:2014sza}. In addition to energies, local chiral interactions describe charge radii extremely well as shown in Fig. 4 of Ref.~\cite{Gandolfi:2020pbj} (see this reference for a more extensive discussion).

\begin{figure}[t]
\centering
\includegraphics[width=4.0in]{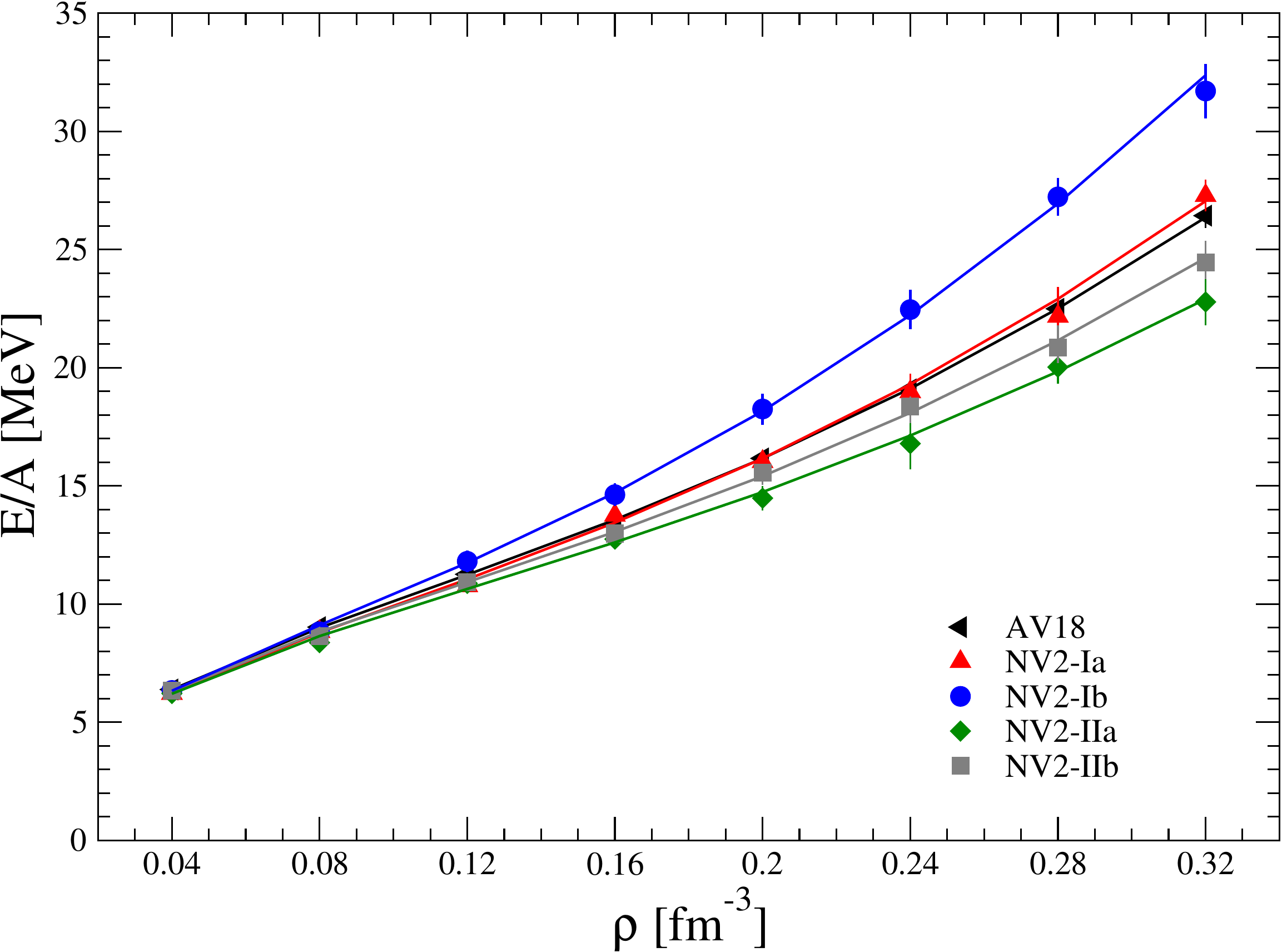}
\caption{From Ref.~\cite{Piarulli:2019pfq}. AFDMC energy per particle of PNM as a function of density for the AV18 (black triangles), NV2-Ia (red triangles), NV2-Ib (solid blue points), NV2-IIa (green diamonds), and NV2-IIb (grey squares) interactions.}
\label{fig:NN_compare}
\end{figure}

The $\Delta$-full models have been recently used in benchmark calculations of the energy per particle of pure neutron matter (PNM) as a function of density using three independent many-body methods~\cite{Piarulli:2019pfq}: Brueckner-Bethe-Goldstone (BBG), Fermi hypernetted chain/single-operator chain (FHNC/SOC), and AFDMC. These calculations are especially useful in providing a quantitative assessment of systematic errors associated with the different many-body approaches and how they depend
on the chosen interaction. A selection of results is reported in Fig.~\ref{fig:NN_compare}, where the energy per particle of pure neutron matter as obtained from AFDMC calculations with the phenomenological AV18 and the NV2 models is reported. The inclusion of $3N$ interactions is essential for a realistic description of neutron matter. Preliminary AFDMC calculations of the equation of state of PNM carried out with the NV2+3-Ia/b and NV2+3-IIa/b models are not compatible with the existence of two solar masses neutron stars, in conflict with recent observations~\cite{Demorest:2010bx,Antoniadis:2013pzd}. On the other hand, the smaller
values of $c_E$ characterizing the $3N$ interactions entering the NV2+3-Ia*/b* and NV2+3-IIa*/b* models mitigate, if not resolve, this problem. There are
indications that these models also predict the energies of low-lying states in light nuclei reasonably well, than 4\% away from the experimental values. Studies along this line are currently in progress.

\begin{acknowledgements}
This research is supported by the U.S. Department of Energy through the FRIB Theory Alliance award DE-SC0013617 (M.P.)
and the U.S. Department of Energy, Office
of Nuclear Science, under contracts DE-AC05-06OR23177 (R.S.).
\end{acknowledgements}


\begin{thebibliography}{}
%
%
\bibitem{Barrett:2013nh}
B.~R.~Barrett, P.~Navratil and J.~P.~Vary,
``Ab initio no core shell model,''
Prog. Part. Nucl. Phys. \textbf{69}, 131-181 (2013).

\bibitem{Jurgenson:2013yya}
E.~D.~Jurgenson, P.~Maris, R.~J.~Furnstahl, P.~Navratil, W.~E.~Ormand and J.~P.~Vary,
``Structure of $p$-shell nuclei using three-nucleon interactions evolved with the similarity renormalization group,''
Phys. Rev. C \textbf{87}, no.5, 054312 (2013).

\bibitem{Hagen:2013yba}
G.~Hagen, T.~Papenbrock, A.~Ekstr\"om, K.~A.~Wendt, G.~Baardsen, S.~Gandolfi, M.~Hjorth-Jensen and C.~J.~Horowitz,
``Coupled-cluster calculations of nucleonic matter,''
Phys. Rev. C \textbf{89}, no.1, 014319 (2014).

\bibitem{Hagen:2013nca}
G.~Hagen, T.~Papenbrock, M.~Hjorth-Jensen and D.~J.~Dean,
``Coupled-cluster computations of atomic nuclei,''
Rept. Prog. Phys. \textbf{77}, no.9, 096302 (2014).

\bibitem{Kievsky:2008es}
A.~Kievsky, S.~Rosati, M.~Viviani, L.~E.~Marcucci and L.~Girlanda,
``A High-precision variational approach to three- and four-nucleon bound and zero-energy scattering states,''
J. Phys. G \textbf{35}, 063101 (2008).

\bibitem{Bogner:2009bt}
S.~K.~Bogner, R.~J.~Furnstahl and A.~Schwenk,
``From low-momentum interactions to nuclear structure,''
Prog. Part. Nucl. Phys. \textbf{65}, 94-147 (2010).

\bibitem{Hergert:2012nb}
H.~Hergert, S.~K.~Bogner, S.~Binder, A.~Calci, J.~Langhammer, R.~Roth and A.~Schwenk,
``In-Medium Similarity Renormalization Group with Chiral Two- Plus Three-Nucleon Interactions,''
Phys. Rev. C \textbf{87}, no.3, 034307 (2013).

\bibitem{Dickhoff:2004xx}
W.~H.~Dickhoff and C.~Barbieri,
``Selfconsistent Green's function method for nuclei and nuclear matter,''
Prog. Part. Nucl. Phys. \textbf{52}, 377-496 (2004).

\bibitem{Soma:2012zd}
V.~Soma, C.~Barbieri and T.~Duguet,
``Ab-initio Gorkov-Green's function calculations of open-shell nuclei,''
Phys. Rev. C \textbf{87}, no.1, 011303 (2013).

\bibitem{Carlson:2014vla}
J.~Carlson, S.~Gandolfi, F.~Pederiva, S.~C.~Pieper, R.~Schiavilla, K.~E.~Schmidt and R.~B.~Wiringa, 
``Quantum Monte Carlo methods for nuclear physics,''
Rev. Mod. Phys. \textbf{87}, 1067 (2015).

\bibitem{Gandolfi:2020pbj}
S.~Gandolfi, D.~Lonardoni, A.~Lovato and M.~Piarulli,
``Atomic nuclei from quantum Monte Carlo calculations with chiral EFT interactions,''
Front. in Phys. \textbf{8}, 117 (2020).

\bibitem{Lahde:2013uqa}
T.~A.~L\"ahde, E.~Epelbaum, H.~Krebs, D.~Lee, U.~G.~Mei\ss{}ner and G.~Rupak,
``Lattice Effective Field Theory for Medium-Mass Nuclei,''
Phys. Lett. B \textbf{732}, 110-115 (2014).

\bibitem{Hagen:2015yea}
G.~Hagen, A.~Ekstr\"om, C.~Forss\'en, G.~R.~Jansen, W.~Nazarewicz, T.~Papenbrock, K.~A.~Wendt, S.~Bacca, N.~Barnea and B.~Carlsson, \textit{et al.}
``Neutron and weak-charge distributions of the $^{48}$Ca nucleus,''
Nature Phys. \textbf{12}, no.2, 186-190 (2015).

\bibitem{Lovato:2014eva}
A.~Lovato, S.~Gandolfi, J.~Carlson, S.~C.~Pieper and R.~Schiavilla,
``Neutral weak current two-body contributions in inclusive scattering from $^{12}$C,''
Phys. Rev. Lett. \textbf{112}, no.18, 182502 (2014).

\bibitem{Lovato:2019fiw}
A.~Lovato, N.~Rocco and R.~Schiavilla,
``Muon capture in nuclei: An ab initio approach based on Green's function Monte Carlo methods,''
Phys. Rev. C \textbf{100}, no.3, 035502 (2019).

\bibitem{Hupin:2014iqa}
G.~Hupin, S.~Quaglioni and P.~Navr\'atil,
``Unified description of $^6$Li structure and deuterium-$^4$He dynamics with chiral two- and three-nucleon forces,''
Phys. Rev. Lett. \textbf{114}, no.21, 212502 (2015).

\bibitem{Elhatisari:2015iga}
S.~Elhatisari, D.~Lee, G.~Rupak, E.~Epelbaum, H.~Krebs, T.~A.~L\"ahde, T.~Luu and U.~G.~Mei\ss{}ner,
``Ab initio alpha-alpha scattering,''
Nature \textbf{528}, 111 (2015).

\bibitem{Fossez:2017wpa}
K.~Fossez, J.~Rotureau, N.~Michel and W.~Nazarewicz,
``Continuum effects in neutron-drip-line oxygen isotopes,''
Phys. Rev. C \textbf{96}, no.2, 024308 (2017).

\bibitem{Fossez:2018gae}
K.~Fossez, J.~Rotureau and W.~Nazarewicz,
``Energy Spectrum of Neutron-Rich Helium Isotopes: Complex Made Simple,''
Phys. Rev. C \textbf{98}, no.6, 061302 (2018).

\bibitem{Inoue:2013nfe}
T.~Inoue \textit{et al.} [HAL QCD],
``Equation of State for Nucleonic Matter and its Quark Mass Dependence from the Nuclear Force in Lattice QCD,''
Phys. Rev. Lett. \textbf{111}, no.11, 112503 (2013).

\bibitem{Beane:2014ora}
S.~R.~Beane, E.~Chang, S.~Cohen, W.~Detmold, H.~W.~Lin, K.~Orginos, A.~Parreno, M.~J.~Savage and B.~C.~Tiburzi,
``Magnetic moments of light nuclei from lattice quantum chromodynamics,''
Phys. Rev. Lett. \textbf{113}, no.25, 252001 (2014).

\bibitem{Orginos:2015aya}
K.~Orginos, A.~Parreno, M.~J.~Savage, S.~R.~Beane, E.~Chang and W.~Detmold,
``Two nucleon systems at $m_\pi\sim 450~{\rm MeV}$ from lattice QCD,''
Phys. Rev. D \textbf{92}, no.11, 114512 (2015)
[erratum: Phys. Rev. D \textbf{102}, no.3, 039903 (2020)].

\bibitem{Savage:2016kon}
M.~J.~Savage, P.~E.~Shanahan, B.~C.~Tiburzi, M.~L.~Wagman, F.~Winter, S.~R.~Beane, E.~Chang, Z.~Davoudi, W.~Detmold and K.~Orginos,
``Proton-Proton Fusion and Tritium $\beta$ Decay from Lattice Quantum Chromodynamics,''
Phys. Rev. Lett. \textbf{119}, no.6, 062002 (2017).

\bibitem{Yukawa:1935xg}
H.~Yukawa,
``On the Interaction of Elementary Particles I,''
Proc. Phys. Math. Soc. Jap. \textbf{17}, 48-57 (1935).

\bibitem{Stoks:1994wp}
V.~G.~J.~Stoks, R.~A.~M.~Klomp, C.~P.~F.~Terheggen and J.~J.~de Swart,
``Construction of high quality N N potential models,''
Phys. Rev. C \textbf{49}, 2950-2962 (1994).

\bibitem{Wiringa:1994wb}
R.~B.~Wiringa, V.~G.~J.~Stoks and R.~Schiavilla,
``An Accurate nucleon-nucleon potential with charge independence breaking,''
Phys. Rev. C \textbf{51}, 38-51 (1995).

\bibitem{Machleidt:2000ge}
R.~Machleidt,
``The High precision, charge dependent Bonn nucleon-nucleon potential (CD-Bonn),''
Phys. Rev. C \textbf{63}, 024001 (2001).

\bibitem{Stoks:1993tb}
V.~G.~J.~Stoks, R.~A.~M.~Klomp, M.~C.~M.~Rentmeester and J.~J.~de Swart,
``Partial wave analaysis of all nucleon-nucleon scattering data below 350-MeV,''
Phys. Rev. C \textbf{48}, 792-815 (1993).

\bibitem{Friar:1984ic}
J.~L.~Friar, B.~F.~Gibson and G.~L.~Payne,
``Recent progress in understanding trinucleon properties,''
Ann. Rev. Nucl. Part. Sci. \textbf{34}, 403-433 (1984).

\bibitem{Pudliner:1997ck}
B.~S.~Pudliner, V.~R.~Pandharipande, J.~Carlson, S.~C.~Pieper and R.~B.~Wiringa,
``Quantum Monte Carlo calculations of nuclei with A \ensuremath{<}= 7,''
Phys. Rev. C \textbf{56}, 1720-1750 (1997).

\bibitem{Navratil:2000gs}
P.~Navratil, J.~P.~Vary and B.~R.~Barrett,
``Large basis ab initio no-core shell model and its application to C-12,''
Phys. Rev. C \textbf{62}, 054311 (2000).

\bibitem{Pieper:2001ap}
S.~C.~Pieper, V.~R.~Pandharipande, R.~B.~Wiringa and J.~Carlson,
``Realistic models of pion exchange three nucleon interactions,''
Phys. Rev. C \textbf{64}, 014001 (2001).

\bibitem{Fujita:1957zz}
J.~Fujita and H.~Miyazawa,
``Pion Theory of Three-Body Forces,''
Prog. Theor. Phys. \textbf{17}, 360-365 (1957).

\bibitem{Pieper:2008rui}
S.~C.~Pieper,
``The Illinois Extension to the Fujita-Miyazawa Three-Nucleon Force,''
AIP Conf. Proc. \textbf{1011}, no.1, 143-152 (2008).

\bibitem{Maris:2013rgq}
P.~Maris, J.~P.~Vary, S.~Gandolfi, J.~Carlson and S.~C.~Pieper,
``Properties of trapped neutrons interacting with realistic nuclear Hamiltonians,''
Phys. Rev. C \textbf{87}, no.5, 054318 (2013).

\bibitem{Weinberg:1990rz}
S.~Weinberg,
``Nuclear forces from chiral Lagrangians,''
Phys. Lett. B \textbf{251}, 288-292 (1990).

\bibitem{Weinberg:1991um}
S.~Weinberg,
``Effective chiral Lagrangians for nucleon - pion interactions and nuclear forces,''
Nucl. Phys. B \textbf{363}, 3-18 (1991).

\bibitem{Weinberg:1992yk}
S.~Weinberg,
``Three body interactions among nucleons and pions,''
Phys. Lett. B \textbf{295}, 114-121 (1992).

\bibitem{Epelbaum:2008ga}
E.~Epelbaum, H.~W.~Hammer and U.~G.~Meissner,
``Modern Theory of Nuclear Forces,''
Rev. Mod. Phys. \textbf{81}, 1773-1825 (2009).

\bibitem{Machleidt:2011zz}
R.~Machleidt and D.~R.~Entem,
``Chiral effective field theory and nuclear forces,''
Phys. Rept. \textbf{503}, 1-75 (2011).

\bibitem{Machleidt:2016rvv}
R.~Machleidt and F.~Sammarruca,
``Chiral EFT based nuclear forces: Achievements and challenges,''
Phys. Scripta \textbf{91}, no.8, 083007 (2016).

\bibitem{Machleidt:2017vls}
R.~Machleidt,
``Historical perspective and future prospects for nuclear interactions,''
Int. J. Mod. Phys. E \textbf{26}, no.11, 1730005 (2017).

\bibitem{Kaplan:1998tg}
D.~B.~Kaplan, M.~J.~Savage and M.~B.~Wise,
``A New expansion for nucleon-nucleon interactions,''
Phys. Lett. B \textbf{424}, 390-396 (1998).

\bibitem{Hammer:2019poc}
H.~W.~Hammer, S.~K\"onig and U.~van Kolck,
``Nuclear effective field theory: status and perspectives,''
Rev. Mod. Phys. \textbf{92}, no.2, 025004 (2020).


\bibitem{Kaplan:1998we}
D.~B.~Kaplan, M.~J.~Savage and M.~B.~Wise,
``Two nucleon systems from effective field theory,''
Nucl. Phys. B \textbf{534}, 329-355 (1998).

\bibitem{Nogga:2005hy}
A.~Nogga, R.~G.~E.~Timmermans and U.~van Kolck,
``Renormalization of one-pion exchange and power counting,''
Phys. Rev. C \textbf{72}, 054006 (2005).

\bibitem{PavonValderrama:2005wv}
M.~Pavon Valderrama and E.~Ruiz Arriola,
``Renormalization of NN interaction with chiral two pion exchange potential. central phases and the deuteron,''
Phys. Rev. C \textbf{74}, 054001 (2006).

\bibitem{Long:2011xw}
B.~Long and C.~J.~Yang,
``Renormalizing Chiral Nuclear Forces: Triplet Channels,''
Phys. Rev. C \textbf{85}, 034002 (2012).

\bibitem{vanKolck:1994yi}
U.~van Kolck,
``Few nucleon forces from chiral Lagrangians,''
Phys. Rev. C \textbf{49}, 2932-2941 (1994).

\bibitem{Rho:1991}
M.\ Rho,
``Exchange currents from chiral Lagrangians,''
Phys.\ Rev.\ Lett.\ {\bf 66}, 1275 (1991).

\bibitem{Furnstahl:2014xsa}
R.~J.~Furnstahl, D.~R.~Phillips and S.~Wesolowski,
``A recipe for EFT uncertainty quantification in nuclear physics,''
J. Phys. G \textbf{42}, no.3, 034028 (2015).

\bibitem{Epelbaum:2014efa}
E.~Epelbaum, H.~Krebs and U.~G.~Mei\ss{}ner,
``Improved chiral nucleon-nucleon potential up to next-to-next-to-next-to-leading order,''
Eur. Phys. J. A \textbf{51}, no.5, 53 (2015).

\bibitem{Furnstahl:2015rha}
R.~J.~Furnstahl, N.~Klco, D.~R.~Phillips and S.~Wesolowski,
``Quantifying truncation errors in effective field theory,''
Phys. Rev. C \textbf{92}, no.2, 024005 (2015).

\bibitem{Wesolowski:2015fqa}
S.~Wesolowski, N.~Klco, R.~J.~Furnstahl, D.~R.~Phillips and A.~Thapaliya,
``Bayesian parameter estimation for effective field theories,''
J. Phys. G \textbf{43}, no.7, 074001 (2016).

\bibitem{Melendez:2017phj}
J.~A.~Melendez, S.~Wesolowski and R.~J.~Furnstahl,
``Bayesian truncation errors in chiral effective field theory: nucleon-nucleon observables,''
Phys. Rev. C \textbf{96}, no.2, 024003 (2017).

\bibitem{Wesolowski:2018lzj}
S.~Wesolowski, R.~J.~Furnstahl, J.~A.~Melendez and D.~R.~Phillips,
``Exploring Bayesian parameter estimation for chiral effective field theory using nucleon\textendash{}nucleon phase shifts,''
J. Phys. G \textbf{46}, no.4, 045102 (2019).

\bibitem{Ordonez:1995rz}
C.~Ordonez, L.~Ray and U.~van Kolck,
``The Two nucleon potential from chiral Lagrangians,''
Phys. Rev. C \textbf{53}, 2086-2105 (1996).

\bibitem{Entem:2003ft}
D.~R.~Entem and R.~Machleidt,
``Accurate charge dependent nucleon nucleon potential at fourth order of chiral perturbation theory,''
Phys. Rev. C \textbf{68}, 041001 (2003).

\bibitem{Epelbaum:2004fk}
E.~Epelbaum, W.~Glockle and U.~G.~Meissner,
``The Two-nucleon system at next-to-next-to-next-to-leading order,''
Nucl. Phys. A \textbf{747}, 362-424 (2005).

\bibitem{Entem:2015xwa}
D.~R.~Entem, N.~Kaiser, R.~Machleidt and Y.~Nosyk,
``Dominant contributions to the nucleon-nucleon interaction at sixth order of chiral perturbation theory,''
Phys. Rev. C \textbf{92}, no.6, 064001 (2015).

\bibitem{Epelbaum:2014sza}
E.~Epelbaum, H.~Krebs and U.~G.~Mei\ss{}ner,
``Precision nucleon-nucleon potential at fifth order in the chiral expansion,''
Phys. Rev. Lett. \textbf{115}, no.12, 122301 (2015).

\bibitem{Reinert:2017usi}
P.~Reinert, H.~Krebs and E.~Epelbaum,
``Semilocal momentum-space regularized chiral two-nucleon potentials up to fifth order,''
Eur. Phys. J. A \textbf{54}, no.5, 86 (2018).

\bibitem{Entem:2017gor}
D.~R.~Entem, R.~Machleidt and Y.~Nosyk,
``High-quality two-nucleon potentials up to fifth order of the chiral expansion,''
Phys. Rev. C \textbf{96}, no.2, 024004 (2017).

\bibitem{Arndt:2007qn}
R.~A.~Arndt, W.~J.~Briscoe, I.~I.~Strakovsky and R.~L.~Workman,
``Updated analysis of NN elastic scattering to 3-GeV,''
Phys. Rev. C \textbf{76}, 025209 (2007).

\bibitem{Perez:2013jpa}
R.~Navarro P\'erez, J.~E.~Amaro and E.~Ruiz Arriola,
``Coarse-grained potential analysis of neutron-proton and proton-proton scattering below the pion production threshold,''
Phys. Rev. C \textbf{88}, no.6, 064002 (2013).

\bibitem{Perez:2013oba}
R.~Navarro P\'erez, J.~E.~Amaro and E.~Ruiz Arriola,
``Coarse grained NN potential with Chiral Two Pion Exchange,''
Phys. Rev. C \textbf{89}, no.2, 024004 (2014).

\bibitem{Perez:2014yla}
R.~Navarro Perez, J.~E.~Amaro and E.~Ruiz Arriola,
``Statistical error analysis for phenomenological nucleon-nucleon potentials,''
Phys. Rev. C \textbf{89}, no.6, 064006 (2014).

\bibitem{Carlsson:2015vda}
B.~D.~Carlsson, A.~Ekstr\"om, C.~Forss\'en, D.~F.~Str\"omberg, G.~R.~Jansen, O.~Lilja, M.~Lindby, B.~A.~Mattsson and K.~A.~Wendt,
``Uncertainty analysis and order-by-order optimization of chiral nuclear interactions,''
Phys. Rev. X \textbf{6}, no.1, 011019 (2016).

\bibitem{Ekstrom:2015rta}
A.~Ekstr\"om, G.~R.~Jansen, K.~A.~Wendt, G.~Hagen, T.~Papenbrock, B.~D.~Carlsson, C.~Forss\'en, M.~Hjorth-Jensen, P.~Navr\'atil and W.~Nazarewicz,
``Accurate nuclear radii and binding energies from a chiral interaction,''
Phys. Rev. C \textbf{91}, no.5, 051301 (2015).

\bibitem{KalantarNayestanaki:2011wz}
N.~Kalantar-Nayestanaki, E.~Epelbaum, J.~G.~Messchendorp and A.~Nogga,
``Signatures of three-nucleon interactions in few-nucleon systems,''
Rept. Prog. Phys. \textbf{75}, 016301 (2012).

\bibitem{Hammer:2012id}
H.~W.~Hammer, A.~Nogga and A.~Schwenk,
``Three-body forces: From cold atoms to nuclei,''
Rev. Mod. Phys. \textbf{85}, 197 (2013).

\bibitem{Hebeler:2020ocj}
K.~Hebeler,
``Three-nucleon forces: Implementation and applications to atomic nuclei and dense matter,''
Phys. Rept. \textbf{890}, 1-116 (2021).

\bibitem{Bernard:2007sp}
V.~Bernard, E.~Epelbaum, H.~Krebs and U.~G.~Meissner,
``Subleading contributions to the chiral three-nucleon force. I. Long-range terms,''
Phys. Rev. C \textbf{77}, 064004 (2008).

\bibitem{Bernard:2011zr}
V.~Bernard, E.~Epelbaum, H.~Krebs and U.~G.~Meissner,
``Subleading contributions to the chiral three-nucleon force II: Short-range terms and relativistic corrections,''
Phys. Rev. C \textbf{84}, 054001 (2011).

\bibitem{Krebs:2012yv}
H.~Krebs, A.~Gasparyan and E.~Epelbaum,
``Chiral three-nucleon force at N$^4$LO I: Longest-range contributions,''
Phys. Rev. C \textbf{85}, 054006 (2012).

\bibitem{Krebs:2013kha}
H.~Krebs, A.~Gasparyan and E.~Epelbaum,
``Chiral three-nucleon force at $N^4LO$ II: Intermediate-range contributions,''
Phys. Rev. C \textbf{87}, no.5, 054007 (2013).

\bibitem{Girlanda:2011fh}
L.~Girlanda, A.~Kievsky and M.~Viviani,
``Subleading contributions to the three-nucleon contact interaction,''
Phys. Rev. C \textbf{84}, no.1, 014001 (2011).

\bibitem{Tews:2015ufa}
I.~Tews, S.~Gandolfi, A.~Gezerlis and A.~Schwenk,
``Quantum Monte Carlo calculations of neutron matter with chiral three-body forces,''
Phys. Rev. C \textbf{93}, no.2, 024305 (2016).

\bibitem{Lynn:2015jua}
J.~E.~Lynn, I.~Tews, J.~Carlson, S.~Gandolfi, A.~Gezerlis, K.~E.~Schmidt and A.~Schwenk,
``Chiral Three-Nucleon Interactions in Light Nuclei, Neutron-$\alpha$ Scattering, and Neutron Matter,''
Phys. Rev. Lett. \textbf{116}, no.6, 062501 (2016).

\bibitem{Lynn:2017fxg}
J.~E.~Lynn, I.~Tews, J.~Carlson, S.~Gandolfi, A.~Gezerlis, K.~E.~Schmidt and A.~Schwenk,
``Quantum Monte Carlo calculations of light nuclei with local chiral two- and three-nucleon interactions,''
Phys. Rev. C \textbf{96}, no.5, 054007 (2017).

\bibitem{Piarulli:2017dwd}
M.~Piarulli, A.~Baroni, L.~Girlanda, A.~Kievsky, A.~Lovato, E.~Lusk, L.~E.~Marcucci, S.~C.~Pieper, R.~Schiavilla and M.~Viviani, \textit{et al.}
``Light-nuclei spectra from chiral dynamics,''
Phys. Rev. Lett. \textbf{120}, no.5, 052503 (2018).

\bibitem{Gazit:2008ma}
D.~Gazit, S.~Quaglioni and P.~Navratil,
``Three-Nucleon Low-Energy Constants from the Consistency of Interactions and Currents in Chiral Effective Field Theory,''
Phys. Rev. Lett. \textbf{103}, 102502 (2009).

\bibitem{Marcucci:2011jm}
L.~E.~Marcucci, A.~Kievsky, S.~Rosati, R.~Schiavilla and M.~Viviani,
``Chiral effective field theory predictions for muon capture on deuteron and $^3He$,''
Phys. Rev. Lett. \textbf{108}, 052502 (2012).

\bibitem{Baroni:2018fdn}
A.~Baroni, R.~Schiavilla, L.~E.~Marcucci, L.~Girlanda, A.~Kievsky, A.~Lovato, S.~Pastore, M.~Piarulli, S.~C.~Pieper and M.~Viviani, \textit{et al.}
``Local chiral interactions, the tritium Gamow-Teller matrix element, and the three-nucleon contact term,''
Phys. Rev. C \textbf{98}, no.4, 044003 (2018).

\bibitem{Schiavilla:2017}
R.~Schiavilla, unpublished (2018).

\bibitem{Wesolowski:2021cni}
S.~Wesolowski, I.~Svensson, A.~Ekstr\"om, C.~Forss\'en, R.~J.~Furnstahl, J.~A.~Melendez and D.~R.~Phillips,
``Fast \& rigorous constraints on chiral three-nucleon forces from few-body observables,''.

\bibitem{Gardestig:2006hj}
A.~Gardestig and D.~R.~Phillips,
``How low-energy weak reactions can constrain three-nucleon forces and the neutron-neutron scattering length,''
Phys. Rev. Lett. \textbf{96}, 232301 (2006).

\bibitem{Gezerlis:2013ipa}
A.~Gezerlis, I.~Tews, E.~Epelbaum, S.~Gandolfi, K.~Hebeler, A.~Nogga and A.~Schwenk,
``Quantum Monte Carlo Calculations with Chiral Effective Field Theory Interactions,''
Phys. Rev. Lett. \textbf{111}, no.3, 032501 (2013).

\bibitem{Gezerlis:2014zia}
A.~Gezerlis, I.~Tews, E.~Epelbaum, M.~Freunek, S.~Gandolfi, K.~Hebeler, A.~Nogga and A.~Schwenk,
``Local chiral effective field theory interactions and quantum Monte Carlo applications,''
Phys. Rev. C \textbf{90}, no.5, 054323 (2014).

\bibitem{Lonardoni:2018nob}
D.~Lonardoni, S.~Gandolfi, J.~E.~Lynn, C.~Petrie, J.~Carlson, K.~E.~Schmidt and A.~Schwenk,
``Auxiliary field diffusion Monte Carlo calculations of light and medium-mass nuclei with local chiral interactions,''
Phys. Rev. C \textbf{97}, no.4, 044318 (2018).

\bibitem{Lynn:2019rdt}
J.~E.~Lynn, I.~Tews, S.~Gandolfi and A.~Lovato,
``Quantum Monte Carlo Methods in Nuclear Physics: Recent Advances,''
Ann. Rev. Nucl. Part. Sci. \textbf{69}, 279-305 (2019).

\bibitem{Piarulli:2016vel}
M.~Piarulli, L.~Girlanda, R.~Schiavilla, A.~Kievsky, A.~Lovato, L.~E.~Marcucci, S.~C.~Pieper, M.~Viviani and R.~B.~Wiringa,
``Local chiral potentials with $\Delta$-intermediate states and the structure of light nuclei,''
Phys. Rev. C \textbf{94}, no.5, 054007 (2016).

\bibitem{Piarulli:2014bda}
M.~Piarulli, L.~Girlanda, R.~Schiavilla, R.~Navarro P\'erez, J.~E.~Amaro and E.~Ruiz Arriola,
``Minimally nonlocal nucleon-nucleon potentials with chiral two-pion exchange including $\Delta$ resonances,''
Phys. Rev. C \textbf{91}, no.2, 024003 (2015).

\bibitem{Piarulli:2020mop}
M.~Piarulli and I.~Tews,
``Local Nucleon-Nucleon and Three-Nucleon Interactions Within Chiral Effective Field Theory,''
Front. in Phys. \textbf{7}, 245 (2020).

\bibitem{Epelbaum:2003gr}
E.~Epelbaum, W.~Gloeckle and U.~G.~Meissner,
``Improving the convergence of the chiral expansion for nuclear forces. 1. Peripheral phases,''
Eur. Phys. J. A \textbf{19}, 125-137 (2004).

\bibitem{Girlanda:2020pqn}
L.~Girlanda, A.~Kievsky, L.~E.~Marcucci and M.~Viviani,
``Unitary ambiguity of NN contact interactions and the 3N force,''
Phys. Rev. C \textbf{102}, 064003 (2020).

\bibitem{Wang:2017}
Meng Wang and G.~Audi and F.G.~Kondev and W.J.~Huang and S.~Naimi and Xing~Xu,
``The {AME}2016 atomic mass evaluation ({II}),''
Chinese Phys. C \textbf{41}, no. 3, 030003 (2017).

\bibitem{Lonardoni:2017hgs}
D.~Lonardoni, J.~Carlson, S.~Gandolfi, J.~E.~Lynn, K.~E.~Schmidt, A.~Schwenk and X.~Wang,
``Properties of nuclei up to $A=16$ using local chiral interactions,''
Phys. Rev. Lett. \textbf{120}, no.12, 122502 (2018).

\bibitem{Piarulli:2019pfq}
M.~Piarulli, I.~Bombaci, D.~Logoteta, A.~Lovato and R.~B.~Wiringa,
``Benchmark calculations of pure neutron matter with realistic nucleon-nucleon interactions,''
Phys. Rev. C \textbf{101}, no.4, 045801 (2020).

\bibitem{Demorest:2010bx}
P.~Demorest, T.~Pennucci, S.~Ransom, M.~Roberts and J.~Hessels,
``Shapiro Delay Measurement of A Two Solar Mass Neutron Star,"
Nature \textbf{467}, 1081-1083 (2010).

\bibitem{Antoniadis:2013pzd}
J.~Antoniadis, P.~C.~C.~Freire, N.~Wex, T.~M.~Tauris, R.~S.~Lynch, M.~H.~van Kerkwijk, M.~Kramer, C.~Bassa, V.~S.~Dhillon and T.~Driebe, \textit{et al.}
Science \textbf{340}, 6131 (2013)

\end{thebibliography}


\end{document}